# The effect of water/carbon interaction strength on interfacial thermal resistance and the surrounding molecular nanolayer of CNT and graphene nanoparticles

Fatemeh Jabbari[1], Ali Rajabpour[2,*], Seyfollah Saedodin[1,†] and Somchai Wongwises[3]

[1] Faculty of Mechanical Engineering, Semnan University, Semnan, Iran

[2] Mechanical Engineering Department, Imam Khomeini International University, Qazvin, Iran

[3] Department of Mechanical Engineering, King Mongkut's University of Technology Thonburi, Bangkok, Thailand

## Abstract

Heat transfer at the liquid/solid interface, especially at the nanoscale, has enormous importance in nanofluids. This study investigates liquid/solid interfacial thermal resistance and structure of the formed molecular nanolayer around a carbon-based nanoparticle. Employing non-equilibrium molecular dynamics simulation and thermal relaxation method, the nanofluid systems with different nanoparticle diameters and different surface wettability were investigated. Simulation results show that carbon nanotubes (CNTs) with a smaller diameter attract more value of the base fluid and lead to a reduced Kapitza resistance. It was found that the thickness of the nanolayer around the nanoparticle is independent of the water/carbon interaction strength. Also, the value of the Kapitza resistance decreases with increasing the interaction strength. Ultimately, a correlation was proposed for the thermal resistance of CNT/water and graphene/water nanofluids in terms of wettability intensity of nanoparticle surface. The proposed correlation in addition to fitting to simulation results can cover the physical conditions of the system.

## 1. Introduction

By diminishing the dimensions of the mechanical and electrical equipment in the industry, the exact examination of the heat transfer mechanism in small sizes is essential, especially at nanoscale. In this way, the study of heat transfer at the joint interface of two different materials has attracted much more attention. Interfacial thermal resistance is known as a criterion for collective resistance between the two

Corresponding authors:
[*] Rajabpour@eng.ikiu.ac.ir
[†] s_sadodin@semnan.ac.ir

elements relative to the flow of heat through them [1]. In 1941, Kapitza [2] was the first who introduced the concept of interfacial thermal resistance (Kapitza resistance). Moreover, liquid layering at the liquid/particle interface is also one of the four factors suggested by Keblinski [3] for the analysis of heat transfer behavior between the nanoparticle and the base fluid. For this reason, many researchers, particularly in the recent years, have studied the properties and structure of this molecular layer in different ways to understand its impact on the physics of the whole system [4-6].

Increasing computers power and continuous improvement of simulation techniques have made the molecular dynamics simulation as a beneficial tool in predicting most of the properties of materials from a model of a statistical physics experiment [7]. The features and advantages of molecular dynamics simulation make it an extremely efficient and flexible tool that can be used to measure thermal properties at the nanoscale and the atomic level [8-10]. For this reason, many researchers have utilized MD simulation to study the thermal properties of systems at the nanoscale such as nanofluids [11-14].

The extraordinary properties of nanofluids compared to conventional fluids have recently led to a significant increase in the volume of studies performed on nanofluids [15-18]. These features can include improved heat transfer and reduced pressure drop, which are the main parameters in the design of electrical and mechanical equipment [19-21]. Among the crucial factors in heat transfer of nanofluid are heat transfer mechanism and particles behavior in the molecular layer of solid/liquid interface. Whatever the dimensions of the system diminish, its effect is significant. Hence, many researchers in recent years have conducted extensive studies on this subject.

Kim et al. [22] studied the heat conduction in nanochannels by equilibrium molecular dynamics (EMD) simulation. Their results indicated a temperature jump and the presence of thermal resistance at the liquid/solid interface. By examining the influence of parameters of the surface wettability and thermal oscillation frequency, they found that by increasing surface wettability and decreasing thermal oscillation frequency, the Kapitza length is reduced, while Xue et al. [23] had already achieved these results. The results of Rajabpour et al. [24] on the Kapitza resistance in hybrid graphene-graphane nanoribbons, which was carried out using non-equilibrium molecular dynamics (NEMD) simulation, revealed that this resistance is highly dependent on the direction of the applied heat flux. In another study, Termentzidis et al. [25] by analyzing the interfacial thermal resistance showed that the high suitability of using rough surfaces instead of soft surfaces reduce the thermal conductivity. Therefore, the importance of heat transfer in the solid/liquid interface has motivated researchers to examine various useful parameters on interfacial thermal resistance; i.e., high frequency and freedom degrees of the two interacting materials [26], intensity of solid/liquid interaction [27], the number of the layers in few-layer graphene [28], interfacial charge decoration [29], overlapping region between solid/liquid [30, 31], the nanoparticle interaction [32] and etc.

In addition to the parameters mentioned above related to the solid/liquid interface, the formation of a molecular layer of liquid around the nanoparticle also affects the heat transfer process. Many studies have discussed the properties and characteristics of this molecular nanolayer to increase the thermal conductivity coefficient. Researchers such as Tillman and Hill [33], Tso et al. [34], Jiang et al. [35, 36], and Milanese et al. [37] studied and computed the effect of molecular nanolayer thickness on thermal conductivity of nanofluid. Cui et al. [38] also examined the impact of parameters such as the shape, size, and material of nanoparticle on the microstructure of molecular nanolayer at the liquid/solid interface. In another study, Guo and Zhao [12] investigated the Cu-Ar nanofluid system by MD simulation. Their results show that the thickness of nanolayer is strongly influenced by the diameter of the nanoparticles, independent of the nanofluid temperature and volume fraction of nanoparticles. Also, Heyhat at al. [39] investigated the Ag-water nanofluid system and studied the effect of the molecular nanolayer at the liquid/solid interface on viscosity and density of nanofluids.

The present research is an attempt to calculate the interfacial thermal resistance and analysis the structure of molecular nanolayer and its thickness at the solid/liquid interface of CNT/water and graphene/water nanofluids and also examine the density distribution of water around the nanoparticles using NEMD simulation and thermal relaxation method. Effects of CNT diameter and wettability of CNT and graphene surfaces on their resistance values are examined. Ultimately, a correlation function is proposed for the interfacial thermal resistance of CNT/water and graphene/water nanofluids regarding wettability intensity of nanoparticle surfaces.

## 2. Interfacial thermal resistance

By passing the heat flux $q''$ from the solid/liquid interface, the thermal resistance between them leads to a temperature jump $\Delta T$ between the two surfaces. This temperature jump occurs for two reasons: a) an incomplete contact between the two surfaces and b) the difference between electrical and vibrational properties of two materials [40]. According to Fourier's law in the heat transfer [29]:

$$q'' = \frac{\Delta T}{R} \quad (1)$$

where $R$ is the interfacial thermal resistance and $G$ is interfacial thermal conductance, which is defined as follows:

$$G = \frac{1}{R} \quad (2)$$

There are two methods for computing the thermal properties of nanofluid such as thermal conductivity and interfacial thermal resistance through MD simulation: equilibrium molecular dynamics (EMD) [41] and non-equilibrium molecular dynamics (NEMD) [4, 8, 9, 24]. According to linear response theory, different thermal properties can be introduced as a form of the Green-Kubo formula. This formula for calculating the interfacial thermal resistance was first proposed by Puech [42] and presented as follows:

$$\frac{1}{R} = G = \frac{1}{Ak_B T^2} \int_0^\infty \langle q(t).q(o) \rangle dt \quad (3)$$

where $A$ is the interface area, $k_B$ is the Boltzmann's constant, $T$ is the system temperature, and $q$ is the oscillating heat power across the interface, which is defined as follows [40]:

$$q(t) = \frac{dE_i(t)}{dt} \quad (4)$$

where $E_i(t)$ is the internal energy of a material at the joint surface at time $t$. The second method is the non-equilibrium molecular dynamics simulation. This method is based on the removal of the system from equilibrium state. In this study, the Kapitza resistance was also measured by applying a solid/liquid temperature gradient, which is discussed in more detail below.

## 3. MD simulation details

### 3.1 Simulation setup

In the present study, water was used as the base fluid in the simulation box with a square cross-section and length of 5.5nm with a CNT at its center. To study the effect of CNT diameter on the Kapitza resistance, the examined systems were further modeled with different diameters while keeping the volume fraction (1.95%) and length (4 nm) of CNT nanoparticles constant. Table 1 provides the characteristics of modeled nanofluid systems to study the effect of CNT diameter.

Table 1. The characteristics of modeled nanofluid systems with constant volume fraction (1.95%) and length (4 nm) of CNT nanoparticles.

| Type of CNT | Diameter of CNT (nm) | Dimensions of the square cross-section in water simulation box (nm) | Number of carbon atoms | Number of water atoms |
|---|---|---|---|---|
| (14,0) | 1.11 | 5.95 | 560 | 18528 |
| (16,0) | 1.27 | 6.8 | 640 | 24243 |
| (18,0) | 1.43 | 7.65 | 720 | 30792 |
| (20,0) | 1.6 | 8.5 | 800 | 37977 |

The initial structure of the CNT/water nanofluid system in two different views is illustrated in Fig. 1.

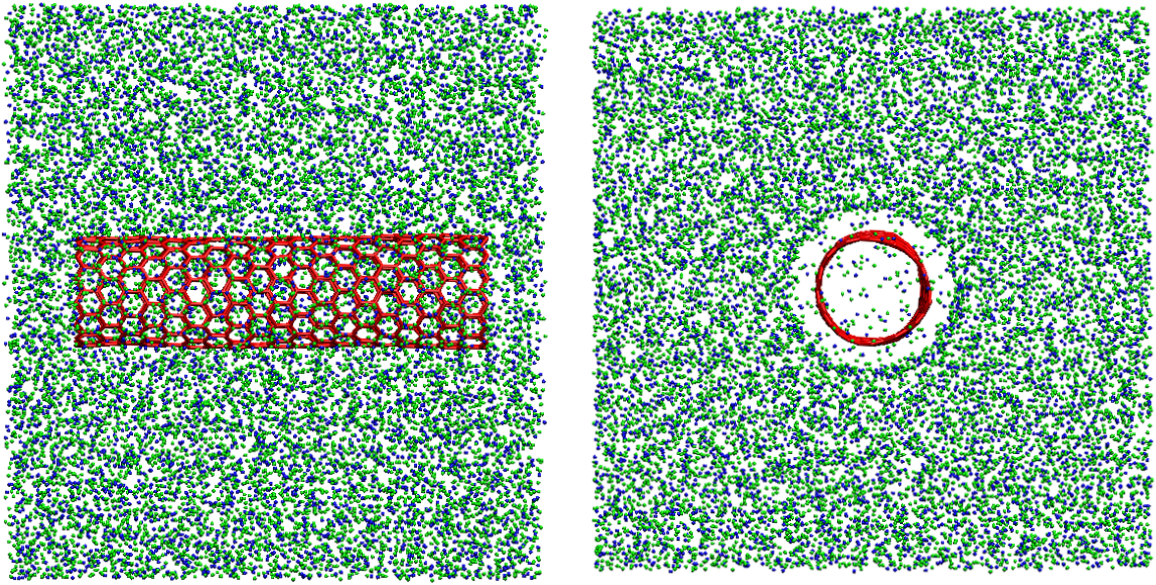

Fig. 1. Nanofluid MD model with periodic boundary conditions in a 3D-computing domain: CNT nanoparticle with the length of 4 nm and surrounding water in the simulation box at a volume fraction of 1.95% in two different views.

Also, to study the Kapitza resistance in graphene/water nanofluid, water was used as the base fluid in the simulation box with a single-layer graphene sheet at its center. The characteristics of the nanofluid systems including graphene are shown in Table 2. Furthermore, Fig. 2 presents the initial structure of the graphene/water nanofluid system in two different views.

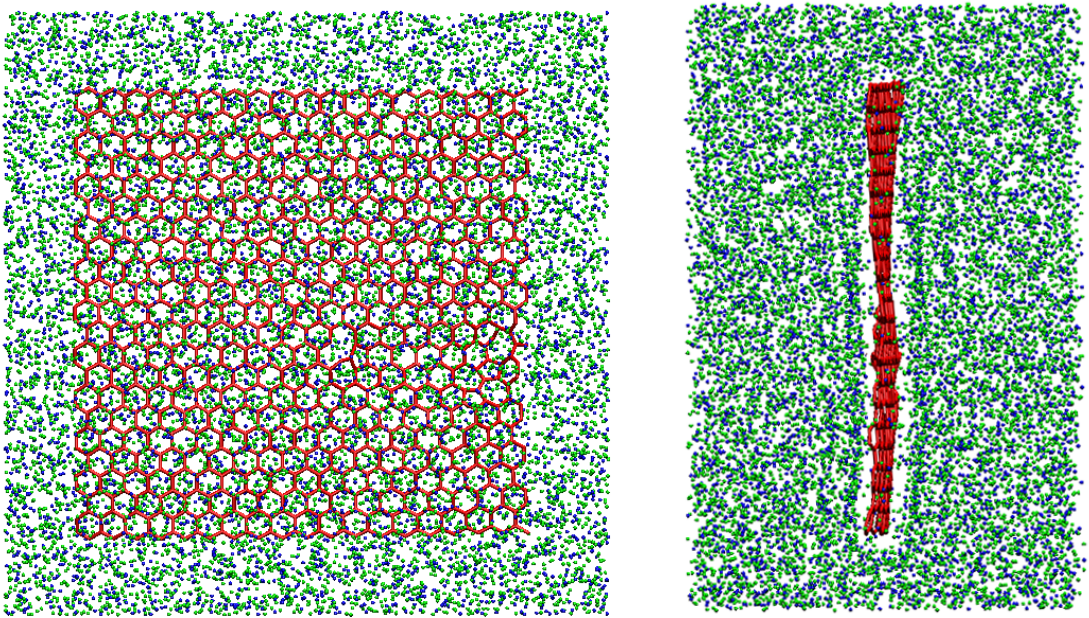

Fig. 2. Nanofluid molecular dynamic model with periodic boundary conditions in a 3D-computing domain: graphene nanoparticle in the water simulation box in two different views.

Table 2. The characteristics of the nanofluid systems including graphene.

| Dimension of graphene (nm) | Dimensions of the water simulation box (nm) | Number of carbon atoms | Number of water atoms |
|---|---|---|---|
| **4.5×4.5** | 6×6×4 | 836 | 14967 |

### 3.2. Simulation processes details

All results of this work were obtained by the large-scale atomic/molecular massively parallel simulator (LAMMPS). The basis calculations of this study are based on the nonequilibrium molecular dynamics simulation. In fact, after bringing the whole system to equilibrium, the temperature gradient is applied between the nanoparticle and the base fluid. Then, the velocity Verlet algorithm is used as an integral model by applying the periodic boundary condition in all three computational directions with a time-step of 0.1 fs. In this simulation, for initial relaxation, the nanofluid system is first placed under the NVE ensemble using the Langevin thermostat over 10 ps. Next, the system is set under the isothermal-isobaric ensemble (NPT) for 50 ps by applying the Nose-Hoover thermostat and barostat [43, 44]. Finally, a canonical ensemble (NVT) is used to the whole system for 10 ps with the Nose-Hoover thermostat to reach the temperature of 300 K and atmospheric pressure. In order to calculate the interfacial thermal resistance, the nanoparticle is placed under the NVT ensemble for 20 ps to reach a temperature of 400 K and then the total system is set under the NVE ensemble to allow heat transfer freely.

The nanofluid system used in this study is composed of single-wall carbon nanotubes (SWCNTs) or graphene and water molecules, while the intermolecular interactions between atoms in the water molecule follow the rigid model of TIP4P/2005. The Lennard-Jones (LJ) potential was chosen for intermolecular interactions between the water molecules with each other and with CNT or graphene. The used LJ parameters in this simulation are listed in Table 3 [10]. Tersoff potential is used as the potential governing the carbon atoms in CNT and graphene nanoparticles.

Table 3. The used LJ parameters in this simulation.

|     | ε(kCal/mol) | σ(Å) |
| --- | --- | --- |
| O-O | 0.185 | 3.16 |
| H-H | 0 | 0 |
| C-O | 0.104 | 3.35 |

### 3.3 Kapitza resistance calculation method

After applying the temperature difference between the nanoparticle and the base fluid, the system is allowed to exchange the heat freely. During this process, the temperature of the nanoparticle and the water at a radius of 1 nm around the nanoparticle (the cut-off distance between carbon atoms and water molecules) and the total energy of the nanoparticle are calculated and stored. Temporal changes in CNT and graphene and water temperature and their difference are shown in Fig. 3 and Fig. 4, respectively.

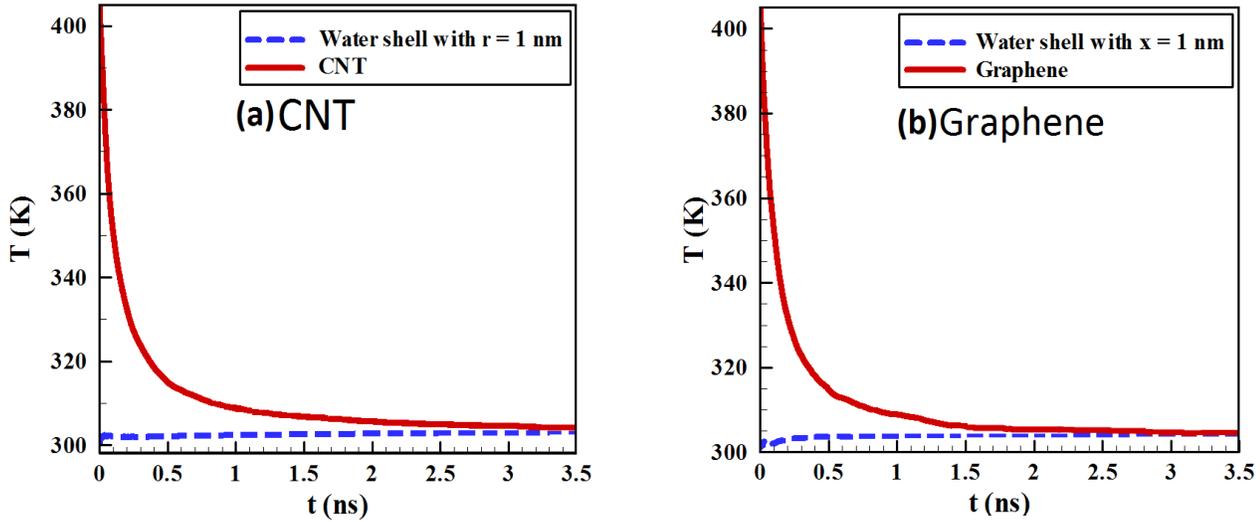
Fig. 3. Temporal changes of nanoparticle and water temperature during the thermal relaxation for a. CNT, b. graphene

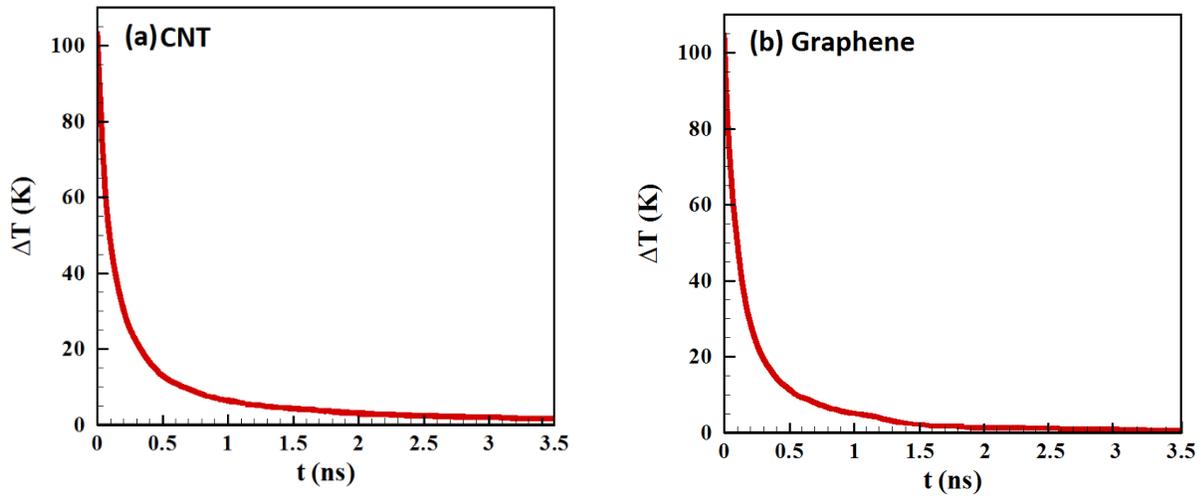
Fig. 4. Temporal changes of the temperature difference between the nanoparticle and water shell with the thickness of 1nm for a. CNT, b. graphene

As expected, the temperature of the nanoparticle and the water converges from their initial value to the equilibrium temperature, until they are thermally balanced as shown in Fig. 4. Also, the changes in the total energy of nanoparticles during the thermal equilibrium time are shown in Fig. 5. Now, by determining the temperature of the water and the nanoparticle and the total energy of the nanoparticle regarding time, the thermal resistance can be calculated according to Eq. (5). This relationship is derived from the the total energy conservation equation for the system.

$$\frac{\partial E_t}{\partial t} = -\frac{T_{NP} - T_W}{R/A} \qquad (5)$$

where $E_t$ is the total energy of the nanoparticle, $A$ is the heat transfer surface, and $R$ is the thermal resistance between water and nanoparticle. By integrating relation (5) and assuming a constant value of R, relation (6) is obtained as follows:

$$E_t = -\frac{A}{R}\int_0^t (T_{NP} - T_W)dt + E_0 \qquad (6)$$

where $E_0$ is the initial total energy of the nanoparticle. To prove that the thermal resistance value is constant, linear variation of $E_t$ versus $\int \Delta T.dt$ is shown in Fig. 6. Now, with the linear fit of this diagram and the calculation of the fitted line slope and the heat transfer surface, it is possible to calculate the thermal resistance between water and nanoparticle.

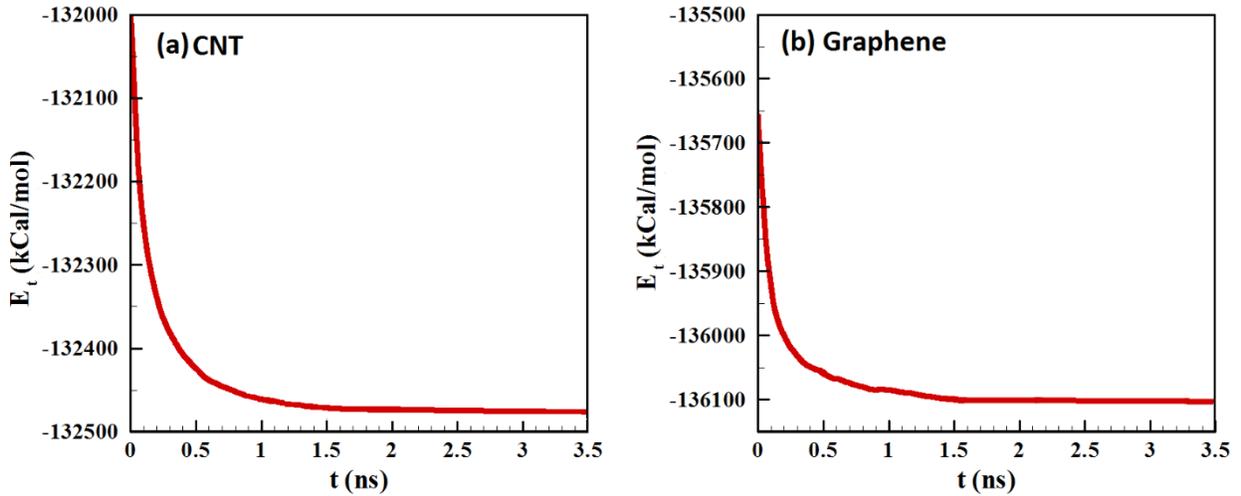

Fig. 5. Temporal changes of the total energy of nanoparticle for a. CNT, b. graphene

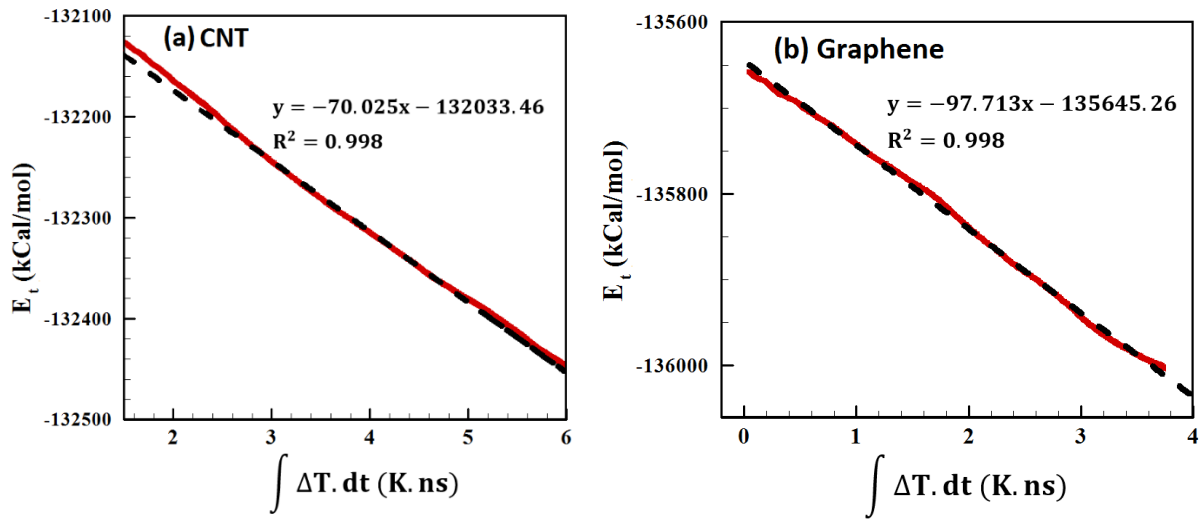

Fig. 6. Linear variation of $E_t$ versus $\int \Delta T.dt$ for a. CNT, b. graphene

## 4. Results and discussion

As previously noted in many studies, the formation of a molecular layer of the base fluid around the nanoparticle and its structure has a significant influence on the behavior of the heat transfer at the solid/liquid interface. In this section, the effect of CNT diameter and wettability of CNT and graphene surfaces on the structure of nanolayer and thermal resistance are examined. A schematic of the division of regions around CNT and graphene are shown in Fig. 7, where the thickness of the shells is 1Å. The density is measured in each of the regions in order to determine the distribution of the water density around the nanoparticle.

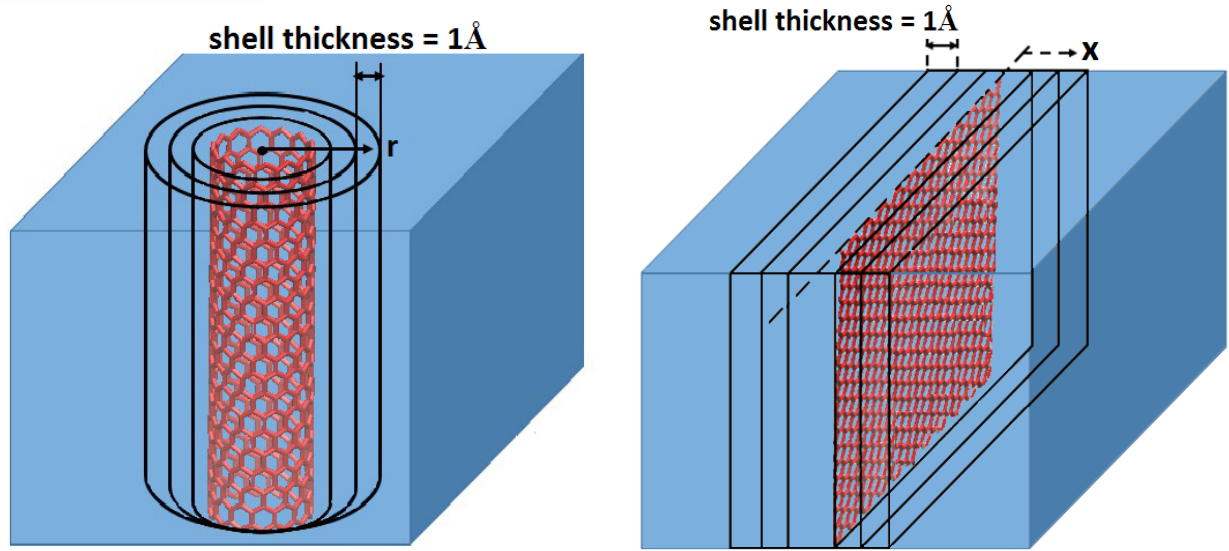

Fig. 7. Schematic of the division of regions around CNT and graphene in order to compute the density profile of water.

### 4.1 Effects of CNT diameter

The variations of the water density versus the distance from the CNT surface for different diameters 11.10, 12.68, 14.27, and 15.86Å are shown in Fig. 8. It can be seen that, due to the absorption of the base fluid by the nanoparticle, the density of water near the nanoparticle is high. As the diameter of the CNT increases, the amount of the first peak curve decreases. In other words, carbon nanotubes with a smaller diameter attract the more value of the base fluid themselves. Thus, the formed nanolayer as a stronger bridge can play the role of heat transfer between the nanoparticle and the base fluid. Finally, it is expected that CNTs with lower diameter have a better ability to transfer heat and therefore have less thermal resistance. Fig. 9 shows the variation of the Kapitza resistance at the solid/liquid interface versus the CNT diameter.

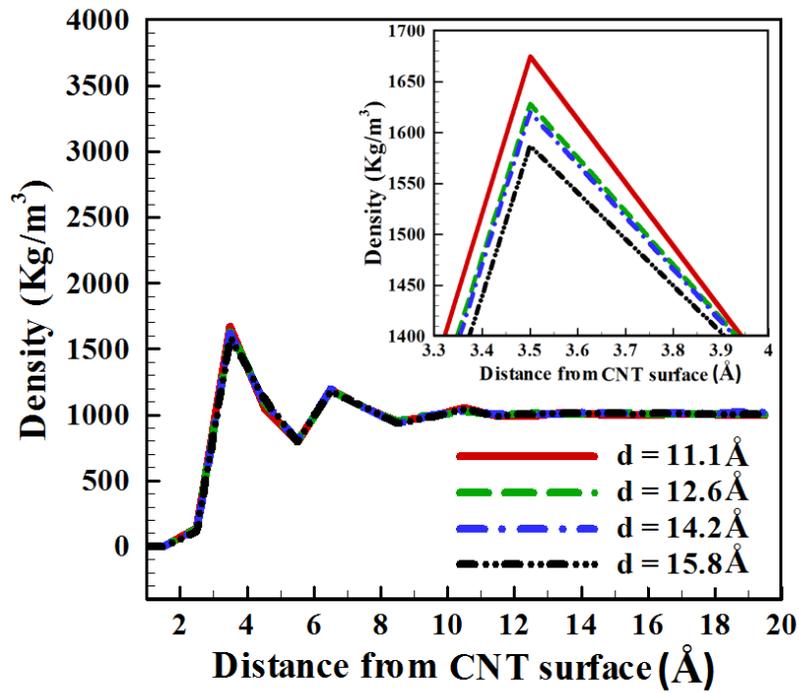

Fig. 8. The variations of the water density around the CNT versus the distance from the CNT surface for different diameters.

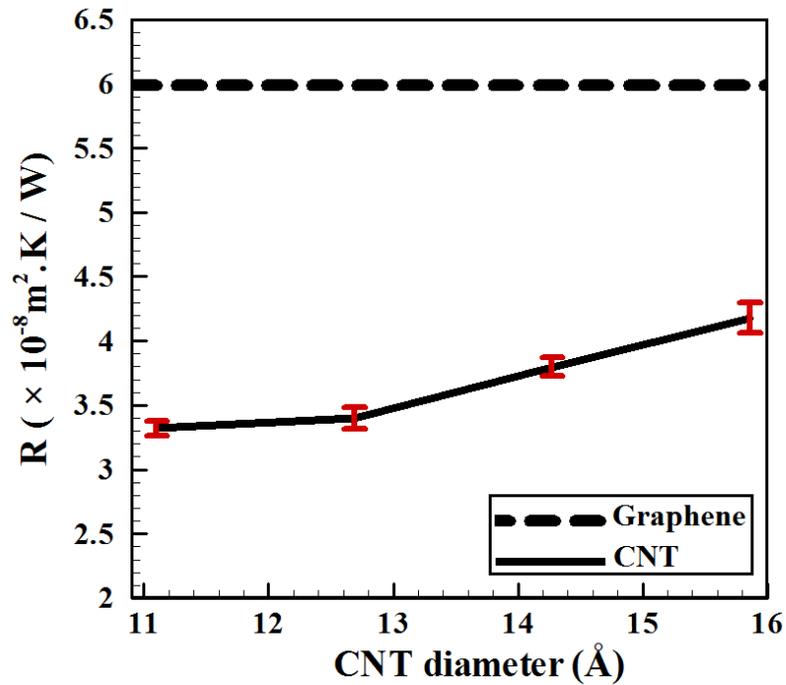

Fig. 9. The variety of the Kapitza resistance at solid/liquid interface versus the CNT diameter.

## 4.2 Effects of wettability of nanoparticle surface

As noted earlier, the interaction between water molecules and carbon atoms follows the Lenard-Jones potential function, where the parameter ε denotes the intensity of the interaction between particles. So, to investigate the effects of wettability of CNT and graphene surfaces, parameter α is considered as follows.

$$\alpha = \frac{\varepsilon_{pf}}{\varepsilon_{ff}} \tag{7}$$

where $\varepsilon_{pf}$ is the energy parameter of Lenard-Jones potential between nanoparticle and base fluid atoms and $\varepsilon_{ff}$ is the energy parameter of Lenard-Jones potential between base fluid atoms. To investigate the effects of wettability of CNT and graphene surfaces on the distribution of the water density around the nanoparticle and the amount of thermal resistance, different values of α equal 0.5, 1, 2, 3, and 4 were considered. Fig. 10 shows the variations of the water density around the nanoparticle versus the different value of α. It can be seen that the interaction strength between water and carbon atom does not have a significant effect on the thickness of the nanolayer around the nanoparticle. In other words, the thickness of the nanolayer is almost constant for different values of wettability of nanoparticle surface. In addition, the value of the first peak of this curve as a powerful factor for heat transfer at the solid/liquid surface is more for stronger interactions. The explanation for this result is that in this case, a better vibrational correlation occurs between the absorbed liquid and the solid surface [27].

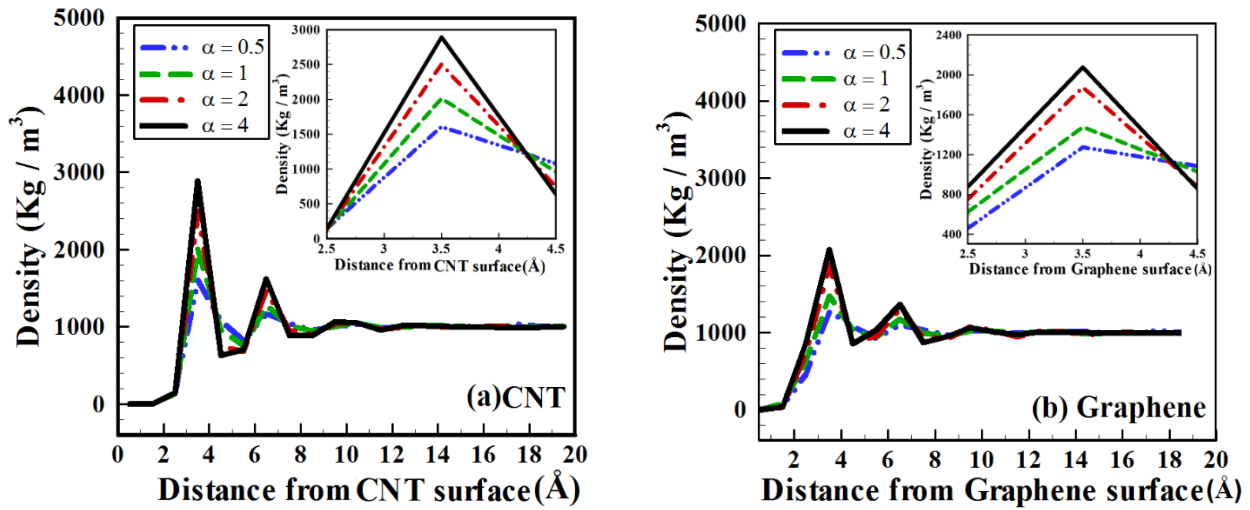

Fig. 10. The variations of the water density around the nanoparticle versus the distance from the nanoparticle surface for different value of α for a. CNT, b. Graphene.

From the analysis of the variations of the water density around the nanoparticle one can expect that by increasing the interaction intensity between the nanoparticle and the base fluid, the amount of heat transfer through the solid/liquid interface increases and consequently the thermal boundary resistance decreases. To support this claim, the variation of the Kapitza resistance at solid/liquid interface versus the wettability of nanoparticle surface is shown in Fig. 11.

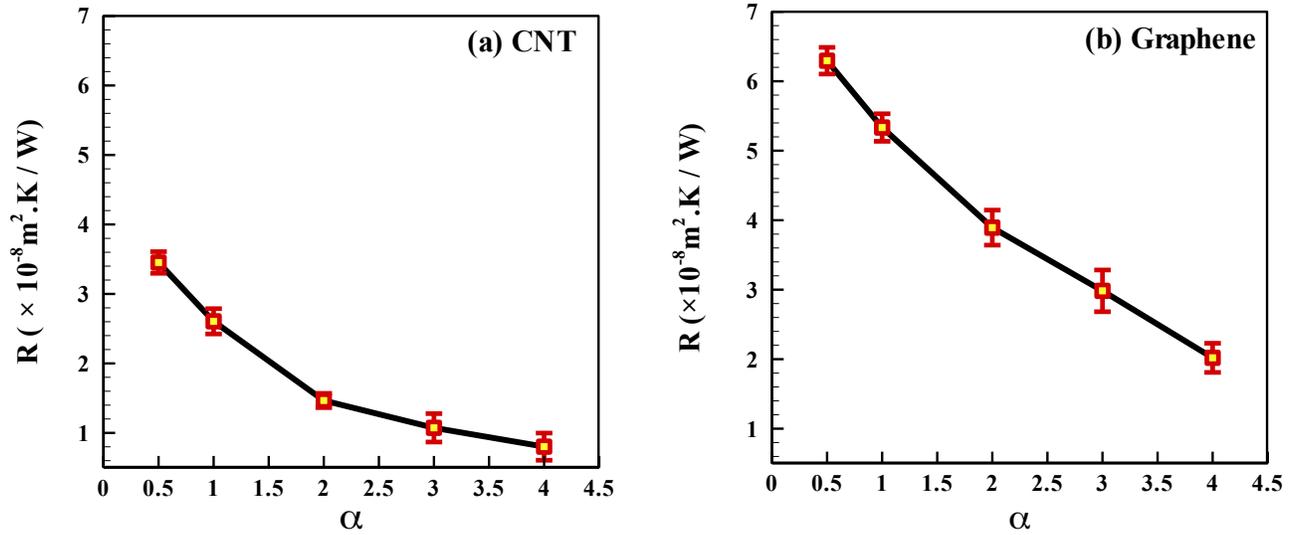

Fig. 11. The variation of the Kapitza resistance at solid/liquid interface versus the wettability of nanoparticle surface for a. CNT, b. Graphene.

### 4.3 Proposed correlation

Since this work studied the influences of the wettability of nanoparticle surface for a limited number of wettability values only, it was desired to develop a relationship to represent the thermal resistance at different wettability of nanoparticle surface. In previous studies [22, 23], researchers have shown that the interfacial thermal resistance shows the power and the exponential behavior for α > 1 and α < 1, respectively. According to the α studied range in this work, a correlation was proposed for the thermal resistance of CNT/water and graphene/water nanofluids in terms of wettability intensity of nanoparticle surface based on the simulation data for CNT and graphene nanoparticles:

$$R = \alpha^a + b\exp(c\alpha) \tag{8}$$

where $a$, $b$, and $c$ are constant values, as presented in Table 4.

Table 4. Constants values of proposed correlations (Eq. 8).

|  | a | b | c |
|---|---|---|---|
| CNT | −0.3062 | 3.283 | −0.7592 |
| Graphene | −0.266 | 6.134 | −0.3538 |

This proposed correlation in addition to displaying very accurate simulation results can cover the physical conditions of the problem. Because with the α value close to zero and infinity, the Kapitza resistance is inclined to infinity and zero, respectively.

To attend the punctuality of the proposed correlation for the nanofluid thermal resistance, the presented correlation predictions were compared to the MD simulation results regarding wettability intensity of nanoparticle surface (Fig. 12).

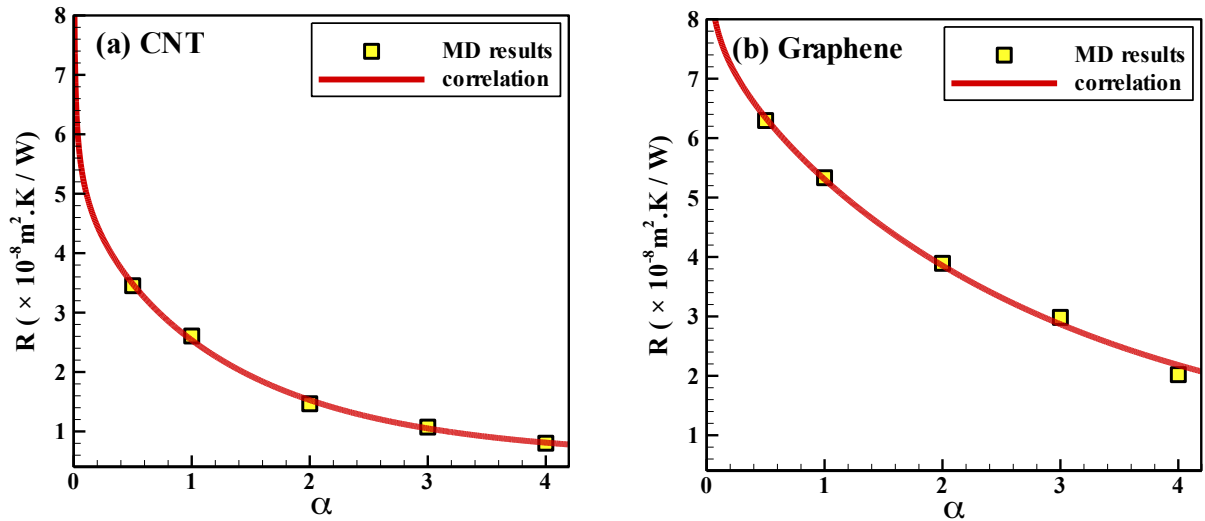

Fig. 12. Comparison between the presented correlation predictions (Eq. 7 and Eq. 8) and MD simulation results for a. CNT, b. Graphene.

## 5. Conclusion

In this study, the effects of CNT diameter and wettability of CNT and graphene surfaces on the interfacial thermal resistance and the microstructure of molecular nanolayer at the solid/liquid interface were examined using the NEMD simulation and thermal relaxation method. Furthermore, a correlation function was proposed for the interfacial resistance of CNT/water and graphene/water regarding wettability of nanoparticle surfaces that by comparing the simulation results with those of the proposed correlation model, the suggested correlation was found to be accurate enough. From the molecular dynamics simulation of this study for CNT/water and graphene/water nanofluid systems, the following results were obtained:

a) CNTs with a smaller diameter attract more base fluid. Therefore, the formed nanolayer as a stronger bridge can play the role of heat transfer between the nanoparticle and the base fluid.
b) CNTs with lower diameter have a better ability to transfer heat and therefore have less interfacial thermal resistance.
c) The interaction strength between water and carbon atoms does not have a significant effect on the thickness of the nanolayer around the nanoparticle. In other words, the thickness of the nanolayer is almost constant for different value of wettability of nanoparticle surface.
d) The value of the first peak for a diagram of the variations of the water density around the nanoparticle versus the distance from the nanoparticle surface is higher for stronger interactions.
e) An increase in the interaction intensity between the nanoparticle and the base fluid results in a decrease in the thermal boundary resistance, while the interfacial thermal resistance at the solid/liquid surface for CNT/water nanofluid is less than that of graphene/water nanofluid.
f) A correlation was proposed for the thermal resistance of CNT/water and graphene/water nanofluids in term of wettability intensity of nanoparticle surface. Comparing the simulation results with those of the proposed correlation model, the suggested correlations were found to be accurate enough.

# References

1. Pop, E., *Energy Dissipation and Transport in Nanoscale Devices.* Nano Research, 2010. **3**(3): p. 147–169.
2. Kapitza, P.L., *Heat Transfer and Superfluidity of Helium II.* Physical Review B, 1941. **60**(4): p. 354-355.
3. Keblinski, P., et al., *Mechanisms of Heat Flow in Suspensions of Nano-Sized Particles (Nanofluids).* International Journal of Heat and Mass Transfer, 2002. **45**(4): p. 855-863.
4. Rajabpour, A., Z. Fan, and S.M.V. Allaei, *Inter-Layer and Intra-Layer Heat Transfer in Bilayer/Monolayer Graphene Van Der Waals Heterostructure: Is There a Kapitza Resistance Analogous?* Applied Physics Letters, 2018. **112**: p. 233104-1-233104-15.
5. Chalopin, Y., et al., *Equilibrium Molecular Dynamics Simulations on Interfacial Phonon Transport.* Annual Review of Heat Transfer, 2014. **17**: p. 147-176.
6. Ahmadi, M.H., et al., *A Review of Thermal Conductivity of Various Nanofluids.* Journal of Molecular Liquids, 2018. **265**: p. 181-188.
7. Mohammadi, K., A. Rajabpour, and M. Ghadiri, *Calibration of Nonlocal Strain Gradient Shell Model for Vibration Analysis of a Cnt Conveying Viscous Fluid Using Molecular Dynamics Simulation.* Computational Materials Science, 2018. **148**: p. 104–115.
8. Rajabpour, A. and S. Volz, *Thermal Boundary Resistance from Mode Energy Relaxation Times: Case Study of Argon-Like Crystals by Molecular Dynamics.* Journal of Applied Physics 2010. **108**: p. 094324-1-094324-8.
9. Farahani, H., et al., *Interfacial Thermal Resistance Between Few-Layer $MoS_2$ and Silica Substrates: A Molecular Dynamics Study.* Computational Materials Science, 2018. **142**: p. 1-6.
10. Jabbari, F., A. Rajabpour, and S. Saedodin, *Viscosity of Carbon Nanotube/Water Nanofluid.* Journal of Thermal Analysis and Calorimetry, 2018. **https://doi.org/10.1007/s10973-018-7458-6**.
11. Jabbari, F., A. Rajabpour, and S. Saedodin, *Thermal Conductivity and Viscosity of Nanofluids: A Review of Recent Molecular Dynamics Studies.* Chemical Engineering Science, 2017. **174**: p. 67–81.
12. Guo, H. and N. Zhao, *Interfacial Layer Simulation and Effect on Cu-Ar Nanofluids Thermal Conductivity Using Molecular Dynamics Method.* Journal of Molecular Liquids, 2018. **259**: p. 40–47.
13. Moghaddam, M.B., E.K. Goharshadi, and F. Moosavi, *Structural and Transport Properties and Solubility Parameter of Graphene/Glycerol Nanofluids: A Molecular Dynamics Simulation Study.* Journal of Molecular Liquids, 2016. **222**: p. 82-87.
14. Taheri, M.H., et al., *Wettability Alterations and Magnetic Field Effects on the Nucleation of Magnetic Nanofluids: A Molecular Dynamics Simulation.* Journal of Molecular Liquids, 2018. **260**: p. 209-220.
15. Das, P.K., *A Review Based on the Effect and Mechanism of Thermal Conductivity of Normal Nanofluids and Hybrid Nanofluids.* Journal of Molecular Liquids, 2017. **240**: p. 420-446.
16. Heyhat, M.M. and A. Irannezhad, *Experimental Investigation on the Competition Between Enhancement of Electrical and Thermal Conductivities in Water-Based Nanofluids.* Journal of Molecular Liquids, 2018. **268**: p. 169-175.
17. Zeroual, S., et al., *Viscosity of Ar-Cu Nanofluids by Molecular Dynamics Simulations: Effects of Nanoparticle Content, Temperature and Potential Interaction.* Journal of Molecular Liquids, 2018. **268**: p. 490-496.
18. Ramezanizadeh, M., et al., *Application of Nanofluids in Thermosyphons: A Review.* Journal of Molecular Liquids, 2018. **272**: p. 395-402.
19. Mikkola, V., et al., *Thermal Properties and Convective Heat Transfer Performance of Solid-Liquid Phase Changing Paraffin Nanofluids.* International Journal of Thermal Sciences, 2017. **117**: p. 163-171.

20. Hemmati-Sarapardeh, A., et al., *On the Evaluation of the Viscosity of Nanofluid Systems: Modeling and Data Assessment.* Renewable and Sustainable Energy Reviews, 2018. **81**(1): p. 313-329.
21. Bigdeli, M.B., et al., *A Review on the Heat and Mass Transfer Phenomena in Nanofluid Coolants with Special Focus on Automotive Applications.* Renewable and Sustainable Energy Reviews, 2016. **60**: p. 1615–1633.
22. Kim, B.H., A. Beskok, and T. Cagin, *Molecular Dynamics Simulations of Thermal Resistance at the Liquid-Solid Interface.* Journal of Chemical Physics, 2008. **129**(74701-1-74701-9).
23. Xue, L., et al., *Two Regimes of Thermal Resistance at a Liquid–Solid Interface.* Journal of Chemical Physics, 2003. **118**(1): p. 337-339.
24. Rajabpour, A., S.M.V. Allaei, and F. Kowsary, *Interface Thermal Resistance and Thermal Rectification in Hybrid Graphene-Graphane Nanoribbons: A Nonequilibrium Molecular Dynamics Study.* Applied Physics Letters, 2011. **99**: p. 051917-1-051917-3.
25. Termentzidis, K., et al., *Thermal Conductivity and Thermal Boundary Resistance of Nanostructures.* Nanoscale Research Letters, 2011. **6**: p. 288-1-288-10.
26. Rajabpour, A. and S. Volz, *Universal Thermal Contact Resistance at High Frequencies.* Physical Review B, 2014. **90**(19): p. 195444-1-195444-4.
27. Pham, A.T., M. Barisik, and B. Kim, *Molecular Dynamics Simulations of Kapitza Length for Argon-Silicon and Water-Silicon Interfaces.* International Journal of Precision Engineering and Manufacturing, 2014. **15**(2): p. 323-329.
28. Alexeev, D., et al., *Kapitza Resistance Between Few-Layer Graphene And Water: Liquid Layering Effects.* Nano Letters, 2015. **15**: p. 5744–5749.
29. Ma, Y., et al., *Ordered Water Layer Induced by Interfacial Charge Decoration Leads to an Ultralow Kapitza Resistance between Graphene and Water.* Carbon, 2018. **135**: p. 263-269.
30. Singh, N., et al., *Analysis of Thermal Interfacial Resistance Between Nanofins and Various Coolants.* International Journal for Computational Methods in Engineering Science and Mechanics, 2011. **12**: p. 254–260.
31. Sarode, A., et al., *A Molecular Dynamics Approach of The Role of Carbon Nanotube Diameter on Thermal Interfacial Resistance Through Vibrational Mismatch Analysis.* International Journal of Thermal Sciences, 2017. **122**: p. 33-38.
32. Han, H., S. Merabia, and F. Müller-Plathe, *Thermal Transport at a Solid–Nanofluid Interface: From Increase of Thermal Resistance Towards a Shift of Rapid Boiling.* Nanoscale, 2017. **9**: p. 8314–8320.
33. Tillman, P. and J.M. Hill, *Determination of Nanolayer Thickness for a Nanofluid.* International Communications in Heat and Mass Transfer, 2007. **34**: p. 399–407.
34. Tso, C.Y., S.C. Fu, and C.Y.H. Chao, *A Semi-Analytical Model for the Thermal Conductivity of Nanofluids and Determination of the Nanolayer Thickness.* International Journal of Heat and Mass Transfer, 2014. **70**: p. 202–214.
35. Jiang, H., et al., *The Role of Interfacial Nanolayer in the Enhanced Thermal Conductivity of Carbon Nanotube-Based Nanofluids.* Applied Physics A, 2015. **118**: p. 197–205.
36. Jiang, H., et al., *Effective Thermal Conductivity of Nanofluids Considering Interfacial Nano-Shells.* Materials Chemistry and Physics, 2014. **148**: p. 195-200.
37. Milanese, M., et al., *An Investigation of Layering Phenomenon at the Liquid–Solid Interface in Cu and CuO Based Nanofluids.* International Journal of Heat and Mass Transfer, 2016. **103**: p. 564–571.
38. Cui, W., et al., *Molecular Dynamics Simulation on the Microstructure of Absorption Layer at the Liquid–Solid Interface in Nanofluids.* International Communications in Heat and Mass Transfer, 2016. **71**: p. 75–85.
39. Heyhata, M.M., et al., *Importance of Nanolayer Formation in Nanofluid Properties: Equilibrium Molecular Dynamic Simulations for Ag-Water Nanofluid.* Journal of Molecular Liquids, 2018. **264**: p. 699-705.

40. Liang, Z. and M. Hu, *Tutorial: Determination of Thermal Boundary Resistance by Molecular Dynamics Simulations.* Journal Of Applied Physics, 2018. **123**: p. 191101-1-191101-16.
41. Rajabpour, A. and S. Volz, *Universal Interfacial Thermal Resistance at High Frequencies* Physical Review B, 2014. **90**: p. 195444-1-195444-4.
42. Puech, L., G. Bonfait, and B. Castaing, *Mobility of the3He solid-liquid interface: Experiment and theory.* Journal of Low Temperature Physics, 1986. **62**(3-4): p. 315–327.
43. Melchionna, S., G. Ciccotti, and B.L. Holian, *Hoover NPT Dynamics for Systems Varying in Shape and Size.* Molecular Physics, 1993. **78**(3): p. 533-544.
44. Plimpton, S., *Fast Parallel Algorithms for Short-Range Molecular Dynamics.* Journal of Computational Physics, 1995. **117**: p. 1-19.